\documentstyle[12pt]{article}
\begin{document}

\title{\Large\bf Possible quantum obstructions to the process of  Abelian conversion}

\author{Ricardo Amorim \thanks {\noindent  amorim@if.ufrj.br } and Ronaldo Thibes \thanks{\noindent thibes@if.ufrj.br} \\
Instituto de F\'{\i}sica\\ 
Universidade Federal do Rio de Janeiro\\ 
RJ 21945-970 - Caixa Postal 68528 - Brasil}
\date{}

\maketitle
\abstract
The  procedure for Abelian conversion of second class constraints due to
Batalin, Fradkin, Fradkina and Tyutin is considered at quantum level, by using the field-antifield formalism. It is argued that quantum effects can obstruct the process.
In this case, Wess-Zumino fields may be introduced in order to restore the lost symmetries.

\vfill
\vspace{1cm}
\newpage

\section{Introduction}

\bigskip
First order  systems with  first and second
class constraints \cite{Dirac}\cite{HT} can be quantized
along several lines. The replacement of Dirac brackets (DB's) by (anti)commutators, which is the cornerstone of canonical quantization,
can only be done for simple systems,   
due to the usual complexity of the DB structure.
Other methods,
as the BRST operatorial quantization
\cite{BRST} 
or its functional counterpart,
the Batalin, Fradkin and Vilkovisky (BFV) procedure \cite{BFV} \cite{FF}, essentially keep the difficulties associated to the
existence of second class constraints \cite{HT}. 
Some years ago, Batalin, Fradkin, Fradkina and Tyutin (BFFT) \cite{BFFT} have introduced 
an algorithm which implements the Abelian conversion of the
second class constraints, by extending in a proper way the  
phase space and also redefining the dynamic variables of the
theory to be converted. This results in a system with a gauge structure with only first class constraints
and a trivial symplectic structure. Its quantization can then be
implemented avoiding DB's and related difficulties.
This route has been extensively followed in recent works \cite{examples},
where in general  first order Hamiltonian systems have their second class
constraints Abelianized by the BFFT method and quantized along 
functional procedures. After integrating over the generalized momenta,
effective Lagrangians are obtained, generating in the configuration space the 
terms responsible for the Abelian conversion of the original second class
sector. In a work by Fujiwara, Igarashi and Kubo \cite{Fuji},
the BFFT procedure is applied in order to convert second class
constraints which satisfy anomalous gauge algebras as a starting point. In a more systematic way, Banerjee, Rothe and Rothe \cite{BRR} have considered a related  problem, by implementing
the BFFT procedure directly at quantum level. These last authors use anomalous commutators \cite{BJL} as the fundamental building blocks for the implementation of the BFFT procedure. In references \cite{Fuji}\cite{BRR}, however, it is not considered the situation
where the process of conversion of second class constraints
introduces in the considered theory a gauge invariance which is 
itself obstructed at quantum level. 
Can the BFFT procedure of conversion present any obstruction due to
quantum effects ? 
Another point that seems to be relevant to be understood is the influence in the BFFT conversion procedure in situations with more general gauge structures,
such as those with open algebras.
\bigskip

To investigate these points we utilize here the tools of the field-antifield 
formalism \cite{HT}\cite{BV}\cite{ZJ}\cite{Troost}\cite{GPS}\cite{DJT}, which naturally takes into account  
systems with general gauge algebras. Once one regularization
procedure is chosen, possible gauge obstructions due to the 
presence of anomalies are also naturally considered in that formalism.
Although the field-antifield formalism is essentially a 
Lagrangian procedure, it is powerful enough to treat also
first order (Hamiltonian) systems \cite{BV-BFV}.
In this work we consider the field-antifield quantization of gauge invariant first order systems that had
their second class constraints converted by the BFFT procedure.
In section {\bf 2} we analyze the situation where only second
class constraints are present in the theory. In section {\bf 3}, this is extended to the mixed case, where also first class
constraints can be originally present.
Section {\bf 4} is devoted to apply the ideas introduced in
the first sections to gauge-fixed chiral electrodynamics . We show in this example that it is necessary to introduce not only BFFT fields, but also
Wess-Zumino fields and antifields \cite{WZ} in order to 
find a local quantum action without obstruction
to its gauge symmetries.
In section {\bf 5} we make some additional comments and conclusions.   
\bigskip

\section{Pure second class constraints}
\bigskip

In order to introduce the ideas in a simpler way, let us first consider the case of a first order system with only second class constraints in a phase-space $P$. For simplicity, we are assuming only discrete bosonic phase-space coordinates 
$y^\mu$, $\mu=1,2,...,2N$, and bosonic second class constraints $\chi_\alpha$,
$\alpha=1,2,...,2n$. The extension to more general situations can be trivially done. Defining the fundamental Poisson brackets (PB's) by 

\begin{equation}
\bigl\{y^\mu,\,y^\nu\bigr\}=f^{\mu\nu}\,,\label{1}
\end{equation}

\bigskip
\noindent where $f^{\mu\nu}$ is an antissymmetric and invertible matrix, the PB between  any two functions $A(y)$ and
$B(y)$ 
on $P$ is given by

\begin{equation}
\bigl\{A,\,B\bigr\}={\partial A\over{\partial y^\mu}}
f^{\mu\nu}{\partial B\over{\partial y^\nu}}\,.\label{2}
\end{equation}

\bigskip
In this way, the first class
Hamiltonian
$H=H(y)$ and the constraints $\chi_\alpha=\chi_\alpha(y)$ satisfy 
the PB structure

\begin{eqnarray}
\bigl\{\chi_\alpha,\chi_\beta\bigr\}&=&\Delta_{\alpha\beta}
\,,\nonumber\\
\bigl\{H,\chi_\alpha\bigr\}&=&
V_\alpha^\beta\chi_\beta\,.\label{3}
\end{eqnarray}

\bigskip Since $\chi_\alpha$ are  second class, 
the constraint
matrix $\Delta_{\alpha\beta}$ is regular. 
\bigskip

The functional quantization of a system like the one appearing in
(\ref{3}) can be done along the lines introduced by Senjanovic \cite{Senj}. The vacuum functional, for instance,
is defined by

\begin{eqnarray}
Z&=&\int[dy^\mu]\vert \det\,f\vert^{-{1\over2}}\delta[\chi_\alpha]
\vert\det\Delta\vert^{1\over 2}\nonumber\\
& &\exp\bigl\{i\int dt[B_\mu \dot y^\mu-
H]\bigr\}
\,.\label{4}
\end{eqnarray}

\bigskip
\noindent We note that in the measure  appears the determinant of
the second class constraint matrix  as well as
the determinant of the
symplectic matrix given by (\ref{1}). As a consequence the measure
becomes invariant under canonical transformations on $P$ \cite{HT}.
In the argument of the exponential,
$B_{\mu}$ is related to $f^{\mu\nu}$ through

\begin{equation}
f_{\mu\nu}={{\partial B_\mu}\over{\partial y^\nu}}-{{\partial B_\nu}\over{\partial y^\mu}}
\label{5}
\end{equation}

\noindent and $f^{\mu\nu}$ is the inverse of  $f_{\mu\nu}$.
\bigskip

\bigskip

To implement the BFFT procedure,
we extend the original phase space $P$ (coordinates $y^\mu$) with BFFT variables $\psi^\alpha$, 
$\,(\alpha=1,2,...,2n)\,$ with PB structure given by
\cite{BFFT}

\begin{equation}
\bigl\{\psi^\alpha,\psi^\beta\bigr\}=\omega^{\alpha\beta}
\,.\label{6}
\end{equation}

\bigskip
\noindent In (\ref{6}), $\omega^{\alpha\beta}$ is 
a constant, antissymmetric and invertible matrix. It follows that in the BFFT extended phase space, the PB between two quantities
$A(y,\psi)$ and $B(y,\psi)$ is given by

\begin{equation}
\bigl\{A,\,B\bigr\}={\partial A\over{\partial y^\mu}}
f^{\mu\nu}{\partial B\over{\partial y^\nu}}+
{\partial A\over{\partial \psi^\alpha}}
\omega^{\alpha\beta}{\partial B\over{\partial \psi^\beta}}
\,,\label{7}
\end{equation}

\bigskip
\noindent as the two sectors of the extended phase space are
assumed to be independent.
\bigskip

The general idea of BFFT is to define new constraints
$\tilde\chi_\alpha = \tilde\chi_\alpha(y,\psi)$ and Hamiltonian 
$\tilde H=\tilde H(y,\psi)$
in such a way that

\begin{eqnarray}
\bigl\{\tilde\chi_\alpha,\tilde\chi_\beta\bigr\}&=&0
\,,\nonumber\\
\bigl\{\tilde H,\tilde\chi_\alpha\bigr\}&=&0
\,.\label{8}
\end{eqnarray}

\bigskip
By requiring that  $\tilde A(y,0)=A(y)$ for any quantity $A$ defined  on $P$ (the unitary
gauge implemented by the choice $\psi^\alpha=0$), 
the original theory is recovered. 
In references \cite{BFFT} it is proved that eqs. (\ref{8}), submitted to the
above condition, always have a power series solution in the BFFT variables,
with coefficients with only $y^\mu$ dependence. 
The second class constraints can be extended to

\begin{equation}
\tilde\chi_\alpha(y,\psi)=\chi_\alpha(y)+X_{\alpha\beta}(y)\psi^\beta
+X_{\alpha\beta\gamma}(y)\psi^\beta\psi^\gamma+\dots
\,.\label{9} 
\end{equation}

\bigskip
The condition that $\tilde\chi_\alpha$ satisfy (\ref{8}) imposes restrictions
in their expansion coefficients. As an  example which will be useful later,
the regular matrices $X_{\alpha\beta}$ must satisfy the identity

\begin{equation}
\label{10}
X_{\alpha\beta}\omega^{\beta\gamma}X_{\delta\gamma}=-\Delta_{\alpha\delta}
\,.\end{equation}

\bigskip
If some quantity $A(y)$ is not a second class constraint, it
can also be extended to $\tilde A(y,\psi)$ in order to be involutive with the converted constraints $\tilde\chi_\alpha$.
BFFT show that in this situation

\begin{equation}
\tilde A(y,\psi)=A(y)- \psi^\alpha\omega_{\alpha\beta}X^{\beta\gamma}\{\chi_\gamma,A\}+...
\label{10a}
\end{equation}

\bigskip
\noindent where the dots represent al least second order corrections to $A(y)$.  
Now it is possible to prove that the first order action

\begin{equation}
\tilde S_0=\int\,dt\,[B_\mu \dot y^\mu+B_\alpha\dot\psi^\alpha+\lambda^\alpha\tilde\chi_\alpha-\tilde H]
\label{11}
\end{equation}

\bigskip
\noindent is invariant under the gauge transformations

\begin{eqnarray}
\delta y^\mu&=&\{y^\mu,\tilde\chi_\alpha\}\epsilon^\alpha\,,\nonumber\\
\delta\psi^\alpha&=&\{\psi^\alpha,\tilde\chi_\beta\}\epsilon^\beta\,,\nonumber\\
\delta\lambda^\alpha&=&-\dot\epsilon^\alpha\,.
\label{12}
\end{eqnarray}

\noindent Close to what occurs in (\ref{5}), in (\ref{12}) $B_\alpha$
is related with the inverse of $\omega^{\alpha\beta}$ through

\begin{equation}
\omega_{\alpha\beta}={{\partial B_\alpha}\over{\partial\psi^\beta}}- 
{{\partial B_\beta}\over{\partial\psi^\alpha}}\,.
\label{13}
\end{equation}

\bigskip
By using some of the above equations, it is not difficult to show that actually 

\begin{equation}
\delta[B_\mu \dot y^\mu+B_\alpha\dot\psi^\alpha+\lambda^\alpha\tilde\chi_\alpha-\tilde H]
={d\over{dt}}\{[B_\mu f^{\mu\nu}{{\partial\tilde\chi_\alpha}\over{\partial y^\nu}}+
B_\beta \omega^{\beta\rho}{{\partial\tilde\chi_\alpha}\over{\partial \psi^\rho}}
+\tilde\chi_\alpha]\epsilon^\alpha\}
\,,\label{14}
\end{equation}

\bigskip
\noindent and consequently we prove that (\ref{11}) is indeed invariant under (\ref{12}), provided boundary terms can be discarded.
\bigskip

As we have already observed, the quantization of the system described by action (\ref{11}) can be done along several different but equivalent lines. 
Under the field-antifield formalism \cite{BV}, it is necessary to introduce the antifields $\phi^*_A=(y^*_\mu,\psi^*_\alpha,\lambda^*_\alpha,c^*_\alpha)$ corresponding respectively to the fields
$\phi^A=(y^\mu, \psi^\alpha, \lambda^\alpha,c^\alpha)$, the ghosts $c^\alpha$ considered
here in equal foot to the previous fields. 
It is then easy to see that the field-antifield action

\begin{equation}
\label{15}
S=S_0+\int\,dt\,[y^*_\mu\{y^\mu,\tilde\chi_\alpha\}c^\alpha+
\psi^*_\beta\{\psi^\beta,\tilde\chi_\alpha\}c^\alpha+\lambda^*_\alpha\dot c^\alpha]
\end{equation}

\bigskip
\noindent satisfy the classical master equation

\begin{equation}
\label{16}
{1\over 2}(S,S)=0\,.
\end{equation}

\bigskip
In the above equation we have introduced the antibracket
 $(X,Y) = {\delta_rX\over
\delta\phi^A} {\delta_lY\over\delta\phi^\ast_A}
- {\delta_rX\over \delta\phi^\ast_A}
  {\delta_lY\over \delta\phi^A}$
 for any two quantities $X$ and $Y$. 
As it is well known, (\ref{16})
contains all the gauge structure associated to the action $S$. To fix the gauge we need to
introduce the trivial pairs $\bar c_\alpha\,,\bar\pi_\alpha$ as new fields,
and the corresponding antifields $\bar{c^{*\alpha}},\bar{\pi^{*\alpha}}$,
as well as a gauge-fixing fermion. It is always possible to 
choose

\begin{equation}
\label{17}
\Psi=\bar c_\alpha\psi^\alpha\,,
\end{equation}

\bigskip
\noindent which implements the unitary gauge, but different
choices are available.
It is also necessary to extend the field-antifield action to

\begin{equation}
\label{18}
S\rightarrow S_\Psi=S+\int\,dt\,\bar\pi_\alpha\bar{c^{*\alpha}}
\,.\end{equation}

\bigskip
\noindent in order to implement the gauge fixing introduced by $\Psi$. The gauge-fixed vacuum functional is now  defined as

\begin{equation}
Z_\Psi=\int[d\phi^A][d\phi^*_A][\det \omega]^{-{1\over2}}
\delta[\phi^*_A - {{\partial\Psi}\over{\partial\phi^A}}]     \exp[i\,S_\Psi]
\,.\label{19}
\end{equation}

\bigskip
In the unitary gauge, we observe that
besides the identifications $\bar c^{*\alpha}=\psi^\alpha,\,\psi^*_\alpha=
\bar c_\alpha$, all the other antifields vanish. With this and the use of
eqs. (\ref{9}-\ref{10}), it is not difficult to see that (\ref{19}) reduces exactly to (\ref{4}), as expected.

\bigskip
A fundamental point to be considered at the quantum level
of any gauge theory is if quantum effects can  obstruct the
gauge symmetry. Under the field-antifield formalism the non obstruction is related with
the independence of the (vacuum) functional with respect to redefinitions of the gauge-fixing fermion $\Psi$. This independence occurs if the classical
field-antifield action $S$ can be replaced by some quantum action $W$
satisfying the so-called quantum master equation

\begin{equation}
\label{20}
<\,{1\over 2}(W,W)\, - \, i\hbar\Delta W\,>_{_{\Psi}}
\,=\,0\,\,,
\end{equation}

\bigskip
\noindent where $<{\cal O}>_\Psi$ means the expected value
of ${\cal O}$ calculated with the use of a specific $\Psi$.
In  expression (\ref{20}) we have introduced the potentially singular operator

\begin{equation}
\Delta \equiv
{\delta_r\over\delta\phi^A}{\delta_l\over\delta\phi^\ast_A}
\,.\end{equation}

\bigskip
If now we expand $W$ in
powers of $\hbar$,

\begin{equation} 
W[\phi^A,\phi^{\ast}_A ] = 
S[\phi^A ,\phi^{\ast}_A ] +
\sum_{p=1}^\infty \hbar^p M_p [\phi^A ,\phi^{\ast}_A ]\,,
\end{equation}

\noindent we can write the quantum master equation (\ref{20}) in loop order.
For the two first terms we have

\begin{eqnarray}
\label{21}(S,S) &=& 0\,,\\
\label{22}
(M_1,S) &=& \,i\, \Delta S\,.
\end{eqnarray}

As expected, the tree approximation gives (\ref{16}). Eq. (\ref{22}) is only formal, since the operator
$\Delta$ must be regularized. If it vanishes when applied on $S$, the quantum action $W$ can be identified with $S$. If its action on $S$ gives a non-trivial result but there exists
some $M_1$ expressed in terms of local fields such that (\ref{22}) is satisfied, gauge symmetries are not obstructed at one loop order. Otherwise, the theory presents anomalies which can be defined by

\begin{equation}
\label{23}
{\cal A }[\,\phi, \phi^\ast \,]\, = \, \Delta S + { i \over  \hbar}
( S , M_1 ) \,=\,a_\alpha\,c^\alpha+\dots\,.
\end{equation}

\bigskip
\noindent It can be shown \cite{Troost} that $a_\alpha$ is
the usual gauge anomaly for closed algebra gauge theories.
So if ${\cal A}$ cannot be set to zero, the process of conversion
is obstructed at quantum level. If one can introduce WZ
fields  in order to restore the lost symmetry, then 
the process is successful, but using more fields than those
originally prescribed by BFFT, which should be equal to the
number of second class constraints originally present in the theory.
In the field-antifield formalism, the introduction of WZ
fields \cite{WZ} are necessary to construct some $M_1$ which is a local 
functional of the extended set of fields in order to have the quantum master equation satisfied if 
true gauge anomalies are found. This kind of procedure
depends on the regularization prescription adopted as well as on the specific
model considered. Further discussions, at this stage, would be only
formal and we reserve section {\bf 4}
to discuss some of these points in the context of
a specific example. 
\bigskip

\section{The mixed case}

\bigskip
In order to generalize the situation treated in section {\bf 2}  we
consider first-order systems that can present from the beginning first class
constraints, say, $\gamma_a(y),\,\, a=1,2,\dots ,m$. Keeping the PB structure
already introduced in (\ref{1}-\ref{2}), such a system in general presents a 
constraint algebra given by \cite{HT}

\begin{eqnarray}
\bigl\{\chi_\alpha,\chi_\beta\bigr\}&=&\Delta_{\alpha\beta}
\,,\nonumber\\
\bigl\{\chi_\alpha,\gamma_b\bigr\}&=& \bar C_{\alpha b}^c\gamma_c+
\bar C_{\alpha b}^\beta\chi_\beta\,,\nonumber\\
\bigl\{\gamma_a,\gamma_b\bigr\}&=& \bar C_{a b}^c\gamma_c+
\bar T_{a b}^{\alpha\beta}\chi_\alpha\chi_\beta\,,\nonumber\\
\bigl\{H,\gamma_a\bigr\}&=& \bar V_a^b\gamma_b+
\bar V_a^{\alpha\beta}\chi_\alpha\chi_\beta\,,\nonumber\\
\bigl\{H,\chi_\alpha\bigr\}&=& \bar V_\alpha^b\gamma_b+
\bar V_\alpha^\beta \chi_\beta\,,\label{25}
\end{eqnarray}

\bigskip
\noindent where $H$ and $\chi_\alpha,\,\alpha=1,2,\dots ,2n$, are respectively the first class Hamiltonian and the second class constraints. Defining the Dirac Brackets between any two
quantities $A$ and $B$ in $P$ by

\begin{equation}
\bigl\{A,B\bigr\}_*=\bigl\{A,B\bigr\}-
\bigl\{A,\chi_\alpha\bigr\}\Delta^{\alpha\beta}\bigl\{\chi_\beta,B\bigr\}\,,
\label {26}\end{equation}

\bigskip
\noindent and choosing gauge-fixing conditions $\Theta_a=0$
such that the matrix $\{\gamma_a,\Theta_b\}$ is non-singular,
the Faddeev-Senjanovic path-integral \cite{Senj}\cite{Faddeev}

\begin{eqnarray}
Z&=&\int[dy^\mu]\vert \det\,f\vert^{-{1\over2}}\delta[\chi_\alpha]\delta[\gamma_a]
\delta[\Theta_a]\vert\det\Delta\vert^{1\over 2}\det\{\gamma_a,\Theta_b\}_*\nonumber\\
& &\exp\bigl\{i\int dt[B_\mu \dot y^\mu-
H]\bigr\}
\,.\label{27}
\end{eqnarray}

\noindent defines the quantization of such a system, provided the
algebra is irreducible. Let us keep the possibility of having
open algebras, this is to say, the consistence of the gauge structure given by  (\ref{25}) demands
the introduction of higher rank structure functions.
This is also associated to the existence of gauge algebras that close only on shell. Open algebras will be considered later  
in this section. As in  section {\bf 2}, we continue assuming 
that the Abelian conversion is implemented with the introduction of the $2n$ variables 
$\psi^\alpha$ that have the same symplectic structure defined
in (\ref{6}) and (\ref{13}). Also  any phase space function $A(y)$ can be properly extended
to a corresponding function $\tilde A(y,\psi)$ submitted to
the condition $\tilde A(y,0)=A(y)$, and
having null PB's ( defined as in (\ref{7})) with any converted
constraint $\tilde \chi_\alpha$, also given by (\ref{9},\ref{10}).
Once this process is implemented, the algebraic structure defined by (\ref{25}) is modified to

\begin{eqnarray}
\bigl\{\tilde\chi_\alpha,\tilde\chi_\beta\bigr\}&=&0\,,\nonumber\\
\bigl\{\tilde\chi_\alpha,\tilde\gamma_a\bigr\}&=&0\,,\nonumber\\
\bigl\{\tilde\gamma_a,\tilde\gamma_b\bigr\}&=& \tilde C_{ab}^c\tilde\gamma_c+
\tilde C_{a b}^{\alpha}\tilde\chi_\alpha\,,\nonumber\\
\bigl\{\tilde H,\tilde\gamma_a\bigr\}&=& \tilde V_a^b\gamma_b+
\tilde V_a^{\alpha}\tilde\chi_\alpha\,,\nonumber\\
\bigl\{\tilde H,\tilde\chi_\alpha \bigr\}&=&0\,,
\label{27b}
\end{eqnarray}

\bigskip
Introducing the compact notation
$\phi^i=(y^\mu,\psi^\alpha)$, $\lambda^A=(\lambda^a,\lambda^\alpha)$,
$\tilde \gamma^A=(\tilde \gamma^a,\tilde\gamma^\alpha)$, $\tilde C^C_{ab}=
(\tilde C^a_{bc}, \tilde C^\alpha_{bc})$ and $\tilde V^B_a=(\tilde V^b_a,
\tilde V^\alpha_b)$, $B_i=(B_\mu, B_\alpha)$ and $\epsilon^A=(\epsilon^a,\epsilon^\alpha)$, we see that
close to what happens to action (\ref{11}), in the mixed case the first order action
\bigskip

\begin{equation}
\tilde S_0=\int\,dt\,[B_A \dot \phi^A+\lambda^A\tilde\gamma_A
-\tilde H]
\label{28}
\end{equation}

\bigskip
\noindent is also invariant under some set of gauge transformations,
now given by

\begin{eqnarray}
\delta\phi^i&=&R^i_A\epsilon^A\,,\nonumber\\
\delta\lambda^A&=&R^A_B\epsilon^B\,,
\label{30}
\end{eqnarray}

\noindent where 

\begin{eqnarray}
R^i_A&=&\{\phi^i,\tilde\gamma_A\}\,,\nonumber\\
R^A_B&=&-\delta^A_B{d\over dt} + \delta^a_B\lambda^b\tilde C^A_{ab}+
\delta^a_B\tilde V^A_a
\,\,.\label{31}
\end{eqnarray}

\bigskip
Now, it is not difficult to see that

\begin{eqnarray}
\left[\delta_{1},\delta_{2}\right]\phi^i &=&\left( R^i_A\tilde C^A_{ab}-\tilde S_{0,A}\{\tilde C^A_{ab},\phi^i\}\right)\epsilon^b_{(1)}\epsilon^a_{(2)}\,,\nonumber\\
\left[\delta_{1},\delta_{2}\right]\lambda^A &=& R^A_B\tilde C^B_{ab} \epsilon^b_{(1)}\epsilon^a_{(2)}\nonumber\\
&+&\left(\{\tilde C^A_{ab},\phi^i\}\tilde S_{0,i}
+(\lambda^C\tilde U^{AB}_{cab}+\tilde V^{AB}_{ab})\tilde S_{0,B}\right)\epsilon^b_{(1)}\epsilon^a_{(2)}
\,,\label{32}
\end{eqnarray}

\bigskip
\noindent where we have defined the second order structure functions through
the relations

\begin{eqnarray}
\label{32}
\tilde U^{AB}_{abc}\tilde\gamma_B&=&\tilde C^d_{ab}\tilde C^A_{cd}+\tilde C^d_{bc}\tilde C^A_{ad}+\tilde C^d_{ca}\tilde C^A_{bd}\,,\nonumber\\
&+&\{\tilde\gamma_a,\tilde C^A_{bc}\}+\{\tilde\gamma_c,\tilde C^A_{ab}\}+\{\tilde\gamma_b,\tilde C^A_{ca}\}\,,\nonumber\\
\tilde V^{AB}_{ab}\tilde\gamma_B&=&
\tilde C^c_{ab}\tilde V^A_c+\tilde C^A_{ac}\tilde V^c_b+\tilde C^A_{cb}\tilde V^c_a\nonumber\\
&+&\{\tilde H,\tilde C^A_{ab}\}+\{\tilde\gamma_a ,\tilde V^A_b\}
-\{\tilde\gamma_b ,\tilde V^A_a\}
\,\,.\end{eqnarray}

\bigskip
\noindent As usual, $\tilde S_{0,i}$ and $\tilde S_{0,A}$ mean the functional variations of
action $\tilde S_0$ with respect to $\phi^i$ and $\lambda^A$. The terms in (\ref{32}) depending on them  
represent trivial gauge transformations \cite{GPS}.
Higher order structure functions are calculated in a similar way,
by imposing consistence of gauge variations with
Jacobi identity.
\bigskip

To quantize such a theory along the lines of the field-antifield formalism, we first introduce the classical field-antifield action

\begin{eqnarray}
\label{33}
S&=&\tilde S_0+\int\,dt[\phi^*_i R^i_A c^A+\lambda^*_A R^A_B c^B+{1\over2}c_A^*\tilde C^A_{ab} c^b c^a\nonumber\\
&+&{1\over2}\lambda^*_A\phi^*_i\{\phi^i,\tilde C^A_{ab}\}c^bc^a+
{1\over4}\lambda^*_A\lambda^*_B\left(\lambda^c\tilde U^{BA}_{cab}+
\tilde V^{BA}_{ab}\right)c^bc^a+\dots]\,,
\end{eqnarray}

\bigskip
\noindent where the dots represent contributions to possible higher rank structure functions. A proper gauge fixing can be implemented by
$\Psi=\bar c_\alpha\Theta^\alpha+\bar c_a\tilde\Theta^a$, where $\tilde\Theta^a$ are related to $\Theta^a$ appearing in (\ref{27}) through
the  process of extension defined for instance in (\ref{10a}). The unitary
gauge is naturally implemented if we choose $\Theta^\alpha=\psi^\alpha$. 
Defining a
non-minimal action through $S_\Psi=S+\int dt\bar\pi_A\bar c^{*A}$,
we write the vacuum functional as
in (\ref{19}), but with the set of fields and antifields consistent with
the present case. By using (\ref{9}-\ref{10a}) and the form assumed for
$\Psi$ for the implementation of the unitary gauge, we can show that the functional analogous to (\ref{19}) reduces to the form (\ref{27}). Now, gauge obstructions
can occur not only in the primitive first class sector, but they also can appear in the process of Abelian conversion of the second class constraints.
The discussion of this situation is parallel to that one done in the end of
section {\bf 2} and will not be repeated here.
At this point it is useful to observe that the field-antifield quatization of
first order systems that present open gauge algebras can be implemented by
using the BFFT procedure without any special restrictions, since 
expression (\ref{33}) gives a well-defined functional.

\bigskip
\section{ An example}
\bigskip

A model where the ideas discussed above can be applied in a simple way is given by the first order action

\begin{equation}
 S_0=\int d^Dx\left[\pi^i\dot A_i-{1\over2}{\pi^i}^2
-{1\over4}F_{ij}^2+i\bar\psi\gamma^\mu D_\mu\psi+\lambda^1(J^0+\partial_i\pi^i)+\lambda^2(\partial_i A^i)\right]
\,,\label{34}
\end{equation}

\bigskip 
\noindent where $\mu,\nu,..=0,1,..,D-1$ and $i,j..=1,2,..,D-1$. $\bar \psi\,,\psi$ and $\gamma^\mu$ are usual Dirac spinors and matrices in
$D$ dimensions. We are here assuming that $D$ is even.  
The Faraday tensor is given by $F_{\mu\nu}=\partial_\mu A_\nu-\partial_\nu A_\mu$, the fermionic (chiral) current is defined through 
$J^\mu={g\over2}\bar\psi\gamma^\mu(1-\gamma_5)\psi$ and the covariant
derivative $D_\mu=\partial_\mu-{ig\over2}(1-\gamma_5)A_\mu$.
Action (\ref{34}) of course represents chiral electrodynamics in D dimensions
in the Coulomb gauge \cite{HT}\cite{Sunder}, where the pair $A_0,\,\pi^0$
has been integrated out. Instead of looking on it from this point of view,
we can just consider action (\ref{34}) as a consistent second class system
which is a good candidate for the process of Abelian conversion. First,
we observe from the symplectic structure of $S$ that (see for instance (\ref{5})) ${{\partial^l}\over{\partial\bar\psi(x)}}    {{\partial^r}\over{\partial\dot\psi(y)}}\left(i\int d^{D-1}z\,\, \bar\psi(z)\gamma^0\dot\psi(z)\right)=
i\gamma^0\delta^{D-1}(x-y)$, where the superscripts $l$ and $r$ mean actions from left and from right. As we are using the metric $\eta=diag(-,+,+,...+)$,
$i\gamma^0$ is itself its inverse and we get directly the bracket structure  
$\{\psi(x),\bar\psi(y)\}=i\gamma^0\delta^{D-1}(x-y)$. For the bosonic sector, similar arguments show that $\{A_i(x),\pi^j(y)\}=\delta^j_i\delta^{D-1}(x-y)$. It is interesting to observe that if we had chosen the chiral covariant derivative to be defined as $D_\mu={1\over2}(1-\gamma_5)(\partial_\mu-igA_\mu)$, 
the symplectic matrix would have no inverse and the bracket structure could not be defined for all the components of $\psi$. With our choice, it is easy to see that the constraints

\begin{eqnarray}
\chi_1&=&J^0+\partial_i\pi^i\nonumber\\
\chi_2&=&\partial_i A^i
\label{35}
\end{eqnarray}

\noindent and the Hamiltonian

\begin{equation}
H=\int d^{D-1}x\left[{1\over2}{\pi^i}^2
+{1\over4}F_{ij}^2-i\bar\psi\gamma^i D_i\psi\right]
\label{35a}
\end{equation}
\bigskip

\noindent form a consistent set of second class constraints and first class Hamiltonian since

\begin{eqnarray}
\label{36}
\{\chi_\alpha(x),H\}&=&0\,,\nonumber\\
\{\chi_\alpha(x),\chi_\beta(y)\}&=&
\epsilon^{\alpha\beta}\nabla^2\delta^{D-1}(x-y)\,.
\end{eqnarray}

\bigskip
\noindent  We are using equal time brackets, $\alpha,\beta=1,2$ 
and  $\epsilon^{12}=-\epsilon_{12}=1$.
\bigskip

To implement the BFFT procedure, we introduce a pair of variables
$\phi^\alpha$ such that $\{\phi^\alpha(x),\phi^\beta(y)\}=\epsilon^{\alpha\beta}
\delta^{D-1}(x-y)$. By choosing conveniently the matrices $X_{\alpha\beta}$
(see (\ref{9})-(\ref{10})), the constraints and the Hamiltonian are converted to

\begin{eqnarray}
\tilde\chi_1&=&J^0+\partial_i\tilde \pi^i\nonumber\\
&=&J^0+\partial_i\pi^i+\nabla^2\phi^1\,,\nonumber\\
\tilde\chi_2&=&\partial_i A^i-\phi^2\,,\nonumber\\
\tilde H&=&H(\tilde\pi^i,A_i,\psi,\bar\psi)
\,,\label{37}
\end{eqnarray}
 
\noindent where we have defined

\begin{equation}
\label{38}
\tilde\pi^i=\pi^i+\partial^i\phi^1
\end{equation}

\bigskip
\noindent and the functional form of $H$ is given in (\ref{35a}), but in (\ref{37}) replacing $\pi^i$ by 
$\tilde\pi^i$. It is trivial now to verify that the constraints $\tilde\chi_\alpha$ satisfy
an Abelian algebra and are involutive with respect to $\tilde H$. As a
consequence, the first order action

\begin{equation}
\tilde S_0=\int d^Dx\left[\pi^i\dot A_i+i\bar\psi\gamma^0\dot\psi+
\phi^2\dot\phi^1-\tilde {\cal H}+\lambda^\alpha\tilde\chi_\alpha\right]
\label{39}
\end{equation} 

\bigskip
\noindent is gauge invariant. Actually, if $\delta y^\mu(x)=\{y^\mu(x),
\int d^{D-1}y\tilde\chi_\alpha(y)\epsilon^\alpha(y)\}$ for any field
$y^\mu$ and $\delta\lambda^\alpha=-\dot\epsilon^\alpha$ for the
multipliers, $\delta\tilde S$ vanishes identically. For convenience,
we observe that

\begin{eqnarray}
\delta A_i&=&-\partial_i\epsilon^1\,,\nonumber\\
\delta \pi^i&=&\partial^i\epsilon^2\,,\nonumber\\
\delta\psi&=&-{ig\over2}(1-\gamma_5)\psi\epsilon^1\,,\nonumber\\
\delta\bar\psi&=&{ig\over2}\bar\psi(1+\gamma_5)\epsilon^1\,,\nonumber\\
\delta\phi^1&=&-\epsilon^2\,,\nonumber\\
\delta\phi^2&=&-\nabla^2\epsilon^2\,,\nonumber\\
\lambda^\alpha&=&-\dot\epsilon^\alpha
\,\,.\label{40}
\end{eqnarray}

It is interesting to note that the quantities $\pi^i$ have non-trivial
transformations, contrarily to what is expected for electrodynamics. Now, if we introduce
the quantities $\tilde \pi^0=\phi^2-\partial_i A^i$, $A_0=\phi^1$ and
$\tilde\lambda^1=\lambda^1-A_0$,
we can rewrite action (\ref{39}) as

\begin{eqnarray}
\tilde S&=&\int d^Dx[\tilde\pi^\mu\dot A_\mu+i\bar\psi\gamma^0\dot\psi
-{1\over2}\tilde\pi^{i2}\nonumber\\
&-&{1\over4}F_{ij}^2+i\bar\psi \gamma^i D_i\psi
+(\tilde\lambda^1-A_0)(\partial_i\tilde\pi^i+J^0)-\lambda^2\tilde\pi^0]\,.
\label{41}
\end{eqnarray} 

\bigskip 
\noindent Not only action (\ref{41}) can be written in terms of $\tilde\pi^\mu$, but also the path integral, since the Jacobian
of the transformation is trivially well defined. Also, from (\ref{38}) and the above definitions, we note that

\begin{eqnarray}
\delta A_0&=&-\epsilon^2\,,\nonumber\\
\delta\tilde\pi^\mu&=&0\,,\nonumber\\
\delta\tilde\lambda^1&=&\epsilon^2-\dot\epsilon^1\,,
\label{42}
\end{eqnarray}

\noindent  the other variations  given by (\ref{40}).
These are just the gauge variations and action of chiral electrodynamics
when written in first order. By looking at (\ref{41}), we observe that $A_\mu$ and $\tilde\pi^\mu$ can be taken as canonical pairs. It is also useful to note that due to definitions of 
$A_0$ and $\tilde\pi^0$, the unitary gauge implemented by $\phi^\alpha=0$ now is expressed by $A_0=\partial_i A^i=0$.

\bigskip
>From what has been discussed above we see that at classical level, the BFFT formalism
was able to reverse the gauge fixation and phase space reduction present in (\ref{34}). At quantum level,
however, the conversion of the constraints (\ref{35}) is obstructed,
since we know that chiral electrodynamics is an anomalous theory.
To investigate this point a bit closer, let us follow the lines discussed in sections {\bf2} and {\bf3}, starting by defining a classical field-antifield action corresponding to (\ref{41}):

\begin{eqnarray}
\label{43}
S_\Psi&=&\tilde S-\int d^Dx[\left(A^{*i}\partial_i+i{g\over2}\psi^*(1-\gamma_5)\psi
+i{g\over2}\bar\psi(1+\gamma_5)\bar\psi^*+\tilde\lambda^*_1\partial_0\right)c^1
\nonumber\\
&+&\left(A^{*0}-\tilde\lambda_1+\lambda^*_2\partial_0\right)c^2- \bar\pi_\alpha\bar c^{*\alpha}]
\end{eqnarray}

\bigskip
\noindent where some proper gauge fixing fermion is assumed. As discussed above, the unitary gauge is here implemented by  $\Psi_{unitary}=\int d^Dx\left(\bar c_1A_0+\bar c_2
\partial_iA^i\right)$, but other choices are available. An interesting choice is given by $\Psi_{covariant}=\int d^Dx\left(\bar c_1\lambda^1+\bar c_2\Theta(A_\mu)\right)$, where $\Theta$ is some unspecified gauge fixing functional . The
choice  given above makes the identifications $\lambda^*_1\equiv\bar c^1$ and $\bar c_1^*\equiv \lambda^1$.  So in (\ref{43}) it will appear the terms
$\int d^Dx\left( \bar\pi_1\lambda^1+\bar c^1\left(\dot c^1-c^2\right)\right)$. The integrations over $\bar \pi_1$ and
$\bar c_1$ implies not only that $\lambda^1$ vanishes but that the ghost $c^2$ must be identified with $\dot c^1$. Integrating over  $\tilde\pi^\mu$ and over $\lambda^1$ makes  action (\ref{43}) be written in
its usual covariant Lagrangian form \cite{HT}\cite{GPS}. All of this can be done without problems since there is no anomaly in the bosonic sector of (chiral) electrodynamics. 
\bigskip

At this stage it is necessary to fix some space-time dimension in order to extract concrete results from the field-antifield machinery. Due to its simplicity, let us consider the case where $D=2$. Action
(\ref{43}) or its partially integrated form than describes  the chiral Schwinger model.
By using a consistent regularization \cite{GPS}\cite{WZ}, it is not difficult to see that

\bigskip
\begin{equation}
\Delta S_\Psi={{ig^2}\over {4\pi}}\int d^2x c^1\left[(1-a)\epsilon^{\mu\nu}-\eta^{\mu\nu}\right]\partial_\mu A_\nu
\label{44}
\end{equation}

\bigskip
\noindent where $a$ is some parameter depending on the regularization and is here assumed to be greater than $1$ \cite{BRR}. We observe that there is no local $M_1$ satisfying (\ref{22}). So the BFFT process of conversion of second class constraints, in this
example, is obstructed.
Following however the procedure introduced by Braga and Montani as well as by Gomis and Par\'{\i}s  in references \cite{WZ}, 
we enlarge furthermore the space of fields and antifields introducing a WZ field
$\theta$ as well as its corresponding antifield $\theta^*$. As
the classical action (\ref{39}) or equivalently (\ref{41}) does  not depend on $\theta$, it is trivially invariant under shifts on it \cite{AD}. So
we can extend the field-antifield action (\ref{43}) to
$\bar S=S_\Psi+\int d^2x \theta^*c^1$ \footnote{We observe that since the field
$\theta$ is absent at classical level, its corresponding antifield could be
introduced in the action multiplied by some indefinite ghost $d$. We choose the quantum action where it appears multiplied by $c^1$ because in this situation the theory can be made anomaly free.}
and introduce the WZ term  

\begin{equation}
\label{45}
M_1=-{1\over{4\pi}}\int d^2x\left\{{{a-1}\over2}\partial_\mu\theta\partial^\mu\theta+
\theta[(a-1)\partial_\mu A^\mu+\epsilon^{\mu\nu}\partial_\mu A_\nu]\right\}
\end{equation}

\bigskip
\noindent such that (\ref{22}) is satisfied. 
Since further terms are identically satisfied if we
define $M_p=0$ for $p$ greater than $1$, we obtain a closed form for $W$ at one loop order.

 Resuming, to convert
the system described by the first order action (\ref{34}) into
a gauge invariant system, the quantum action $W=S+\hbar M_1$
had to be extended not only with the aid of the two expected
BFFT fields $\phi^\alpha$, but also with a pair of WZ field and antifield which had  origin in quantum obstructions of the
gauge symmetry classically introduced with the aid of the    BFFT fields $\phi^\alpha$.
\bigskip

\section{Conclusions}  

\bigskip  
We have considered the implementation of the BFFT procedure for converting
first order systems with first and second class constraints at quantum level
in a general way,
by using the field-antifield formalism. We argue that this process can be 
obstructed due to the occurrence of gauge anomalies. When this is the case,
the introduction of further auxiliary (WZ) fields may be considered.
We also have shown  that open gauge algebras play no special role in the process of Abelian conversion, being considered in the usual way under the field-antifield formalism. An example based on quantum chiral electrodynamics has been included.
Specific results have been presented for the case where D=2.
Presently we are studying other models where second class constraints may appear
in a somehow more fundamental way. Also the cohomological version  of these procedures
is under study.  Results will be reported elsewhere.

\vskip 1cm
\noindent {\bf Acknowledgment:} One of the authors (R. A.) is in debt to
R. Banerjee, J. Barcelos-Neto, N. R. F. Braga and M. Henneaux for useful discussions and
comments. 
This work is supported in part by
Conselho Nacional de Desenvolvimento Cient\'{\i}fico e Tecnol\'ogico
- CNPq (Brazilian Research Agency).

\end{document}